# A simulation of the cluster structures in Ge-Se vitreous chalcogenide semiconductors

V. Gurin

*Research Institute for Physical Chemical Problems, Belarusian State University*

*Leningradskaya 14, 220030 Minsk, Belarus*

O. Shpotyuk, V. Boyko

*Institute of Materials of Scientific Research Company "Carat",*

*Stryjska str., 202, 79031 Lviv, Ukraine*

A structure of germanium selenide glasses is simulated by the featured clusters built from the tetrahedral $GeSe_4$ units up to the clusters with six germanium atoms ($Ge_6Se_{16}H_4$ and $Ge_6Se_{16}H_8$). Quantum chemical calculations at the DFT level with effective core potentials for Ge and Se atoms for the clusters of different composition reveal their relative stability and optical properties.

1. **Introduction**

Vitreous chalcogenide semiconductors (VCS) of binary Ge-Se family, are of continuous interest since 60-70[th] of the last century as perspective materials for application in telecommunication, optoelectronics, sensorics, etc. as well as glassy solid media with complicated structure and challenged properties [1]. There is no complete understanding of structural features in VCS to date. One of important phenomena inherent to this type of glassy materials is their ability to self-organization [2-4] that consists in specific interplay between composition and properties. It signs to an existence of unique structure units and subunits providing the highest non-ageing ability, thermal and mechanical stability. The latter is evident requirement for application, but empirical approach is not enough to realize these phenomena. There are many models to reproduce both structural features and physical properties of VCS, however no final solutions at present. A limited disorder in the structure of these glasses creates troubles in direct experimental studies. An atomic-scale simulation of the glass structures with modern quantum chemical methods is an advanced approach that allows get information at *ab initio* or semiempirical levels covering a broad size range of building elements: molecules, clusters, periodic and quasiperiodic structures. In the present work, a quantum chemical simulation of the clusters within Ge-Se VCS is performed at the level of density functional theory (DFT) and feasible structures of medium-range building units are established.

## 2. Calculation details and model building

The calculations were done by DFT method with B3LYP functional using LANL2DZ basis sets with effective core potential (ECP) for heavy atoms (Ge and Se) and the basis 6-311G for H atoms. This ECP makes frozen internal electronic shells of Ge and Se that can respond quite good accuracy for *p*-elements. B3LYP functional is known to be well balanced choice in calculations both molecules and clusters. NWCHEM package was utilized for geometry optimization, and the electronic transitions were calculated with GAUSSIAN03 within the framework of the time-dependent DFT (TDDFT) method (including excited states of clusters covering the energy range up to 5-6 eV).

The clusters in Ge-Se VCS can be built taking $GeSe_4$ tetrahedrons as primary structure units linked together through several possible ways. In this series of calculations we take only one way to join the tetrahedra, edge-sharing, resulting in $Ge_2Se_6$ units (see Fig. 1). The next step is a linking of more $GeSe_4$ tetrahedra to the terminal Se atoms (with extraction of Se atoms providing the heterogeneous chains ≡Ge-Se-Ge≡ in this construction) (Fig. 1). Thus a cluster with $Ge_6Se_{16}$ composition is constructed. In the calculations, it was added with eight terminating hydrogen atoms to saturate dangling bonds and compensate an extra charge, so the final $Ge_6Se_{16}H_8$ cluster is uncharged. A somewhat other composition is obtained assuming an occurrence of bridge -Se-Se- groups (Fig. 2) with 4 terminating H atoms, $Ge_6Se_{16}H_4$.

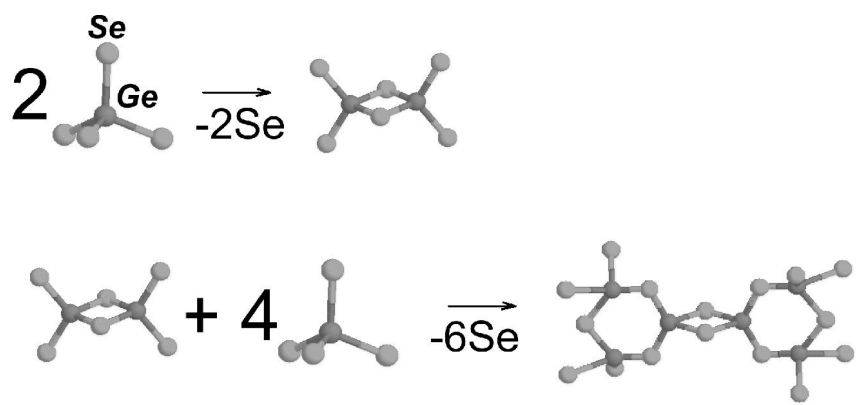

Fig. 1. A scheme of $Ge_6Se_{16}$ model cluster construction: $GeSe_4 \rightarrow Ge_2Se_6 \rightarrow Ge_6Se_{16}$

## 3. Results and discussion

The data collected in Table 1 present energetic and geometric characteristics of two types of Ge-Se clusters built from six basic tetrahedral units (Fig. 2). Binding energy ($E_b$) was estimated with respect to complete cluster destroy per one Ge-Se bond. The values indicate that clusters are rather stable and stability of both clusters is almost the same, i.e. the occurrence of -Se-Se- bridges only lightly diminish $E_b$ in the case of $Ge_6Se_{16}H_4$ cluster. Easy transformations between these two clusters may proceed in a glass structure, and also more number of bridge-type groups is expected to be formed that is familiar feature of these VCS. HOMO-LUMO gaps for both clusters are about 3 eV. This value is more that the bulk energy gap $E_g$ of $GeSe_2$ (2.5 eV) in accordance with expectable variation of band gaps for quantum-sized particles. Analysis of the data for interatomic distances (R) evidences that the cluster without -Se-Se- bridges, $Ge_6Se_{16}H_8$, appears as slightly more expanded than $Ge_6Se_{16}H_4$.

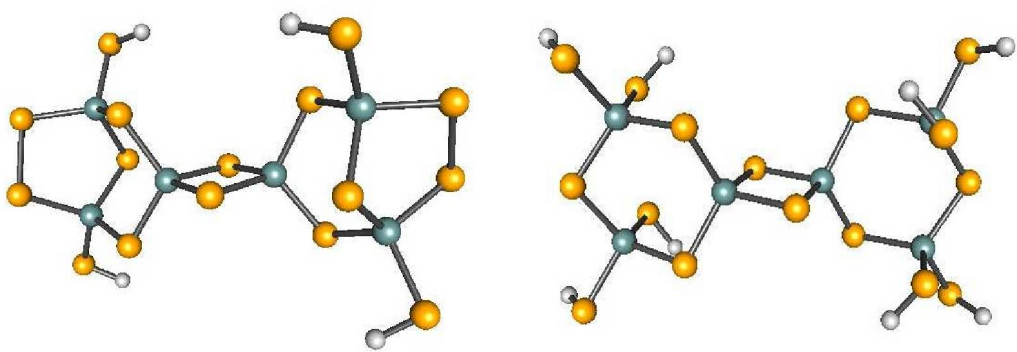

Figure 2. Geometry of the clusters calculated $Ge_6Se_{16}H_4$ (left) and $Ge_6Se_{16}H_8$ (right).

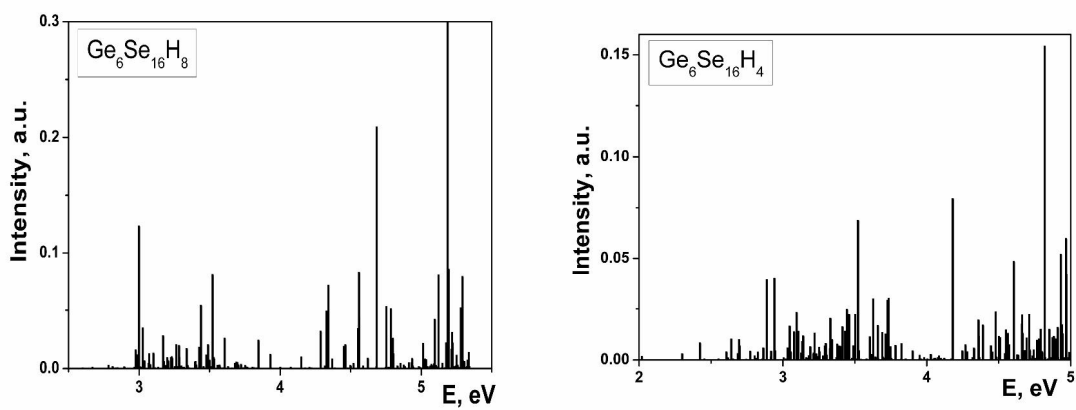

Figure 3. Simulation of optical absorption spectra for the clusters $Ge_6Se_{16}H_4$ (left) and $Ge_6Se_{16}H_8$ (right) through the electronic transitions to excited states.

Table 1. Numerical data for the clusters $Ge_6Se_{16}H_4$ and $Ge_6Se_{16}H_8$

| Cluster | $E_{total}$ (optimized ground state), a.u. | $E_b(eV) = (E_{total} - 6E_{Ge} - 16E_{Se} - _HE_H)/24$ | Energy of HOMO, LUMO, a.u. $\Delta E$, eV | $R_{Ge-Se}$ Å | $R_{Ge-Ge}$ in centre, Å |
|---|---|---|---|---|---|
| $Ge_6Se_{16}H_4$ | -173.1033 | 3.823 | -0.2497 -0.1423 2.923 | 2.486 2.502 2.478 2.464 2.450 2.475 2.448 2.500 | 3.624 |
| $Ge_6Se_{16}H_8$ | -175.4728 | 4.250 | -0.2601 -0.1362 3.372 | 2.507 2.481 2.482 2.460 2.471 2.468 | 3.957 |

In order to determine optical properties of the clusters, a series of electronic transitions were calculated (Fig. 3). The major bands in the UV-Vis range are about 2.9, 3.5, 4.2, and 4.8 eV for the first cluster (with -Se-Se- bridge) and 3.0, 4.7, 5.2 eV for the second one. Thus these model clusters can be used to distinguish an appearance of bridge-like structures within studied Ge-Se VCS. The spectra reveal similarity with excitonic absorption (both experimental and theoretical) of semiconductor nanocrystalline species like CdS, ZnSe, etc. [5-7]. Therefore, the clusters under study simulate fragments of bulk structures in binary Ge-Se glasses which are similar with semiconductor nanoclusters in the formation of excitonic states.

## 4. Conclusion

Clusters of the composition $Ge_6Se_{16}H_4$ and $Ge_6Se_{16}H_8$ were constructed for simulation of featured structures in Ge-Se VCS. Calculations of optimized geometry, energies and electronic transitions were done at the DFT and TDDFT levels. They showed the relative stability of the clusters and simulated optical spectra. So these clusters can be considered as competitive model units for basic structures possible in binary Ge-Se VCS.

**Acknowledgments**

The authors acknowledge support from Belarusian Republican Foundation for Fundamental Research (Project F11K-131) and State Found for Fundamental Research of Ukraine (Project F41.1/044).

**References**

1  N.F. Mott and E.A. Davis, *Electronic Processes in Non-Crystalline Materials* (Clarendon Press, Oxford, 1971).
2  R. Golovchak, O. Shpotyuk, S. Kozyukhin, et al., *J. Appl. Phys.* **105**, 103704 (2009).
3  O. Shpotyuk, R. Golovchak, *Phys. Stat. Solidi* **C8**, 2572 (2011).
4  O. Shpotyuk, R. Golovchak, *J. Optoelectron. Advanced Mater.* **14**, 596 (2012).
5  V.S. Gurin, *Colloids and Surfaces. A: Physicochemical and Engineering Aspects*, **202**, 215 (2002).
6  V.S. Gurin, *Materials Science and Engineering B*, **169**, 73 (2010).
7  V.S. Gurin, *Solid State Communication*, **112**, 631 (1999).